\documentstyle[aps,twocolumn]{revtex}

\def\h{h}

\def\d{d}
\def\p{\partial}
\def\w{\wedge}
\def\o{\otimes}
\def\g{g}
\def\gu{g^{-1}}
\def\h{h}

\def\ric{Ric}
\def\T{T}
\def\j{j}
\def\ix{\iota_{\xi}}
\def\Lx{{\cal L}_{\xi}}
\def\F{F}
\def\A{A}
\def\u{u}

\def\tanh{{\rm tanh}}

\def\Arcsin{{\rm Arcsin}}
\def\R{R}
\def\half{\frac{1}{2}}
\def\quota{\frac{1}{4}}

\def\tr{\tilde{r}}
\def\tt{\tilde{t}}
\def\tphi{\tilde{\phi}}
\def\teta{\tilde{\eta}}

\def\br{\bar{r}}
\def\bphi{\bar{\phi}}
\def\bfeta{\bar{\eta}}
\begin{document}
\twocolumn[\hsize\textwidth\columnwidth\hsize\csname @twocolumnfalse\endcsname
\title{Static charged perfect fluid with Weyl-Majumdar relation}
\author{Daisuke Ida}
\address{
Department of Physics, Kyoto University, Kitashirakawa,
Sakyo-ku, Kyoto 606-8502, Japan\\ 
{\rm E-mail address: ida@tap.scphys.kyoto-u.ac.jp}\\
}
\date{\today}
\maketitle
\begin{abstract}
Static charged perfect fluid distributions have been studied.
It is shown that if the norm of the timelike Killing vector and
the electrostatic potential have the Weyl-Majumdar relation, then
the background spatial metric is the space of constant curvature,
and the field equations reduces to a single non-linear
partial differential equation. Furthermore, if the linear equation of
state for the fluid is assumed, then this equation becomes a
Helmholtz equation on the space of constant curvature. 
Some explicit solutions are given.
\end{abstract}
]
\section{Introduction}
In Newtonian theory, there are static equilibrium configurations of
charged particles 
in which the gravitational attractions and the electric
repulsions among the particles exactly balance.
In general relativity, it is well known that the Einstein equation
has such a static solution corresponding to these Newtonian situations,
namely the Majumdar-Papapetrou solution \cite{Majumdar47,Papapetrou47} describing an equilibrium
state of many charged black holes. \cite{Hartle&Hawking}

Majumdar \cite{Majumdar47} studied the static
Einstein-Maxwell system that is described by the static
metric
\begin{equation}
\g=f\d t\o\d t-f^{-1}\h \label{static}
\end{equation}
and the vector potential
\begin{equation}
\A=-\psi\d t,
\end{equation}
subject to the coupled Einstein and source-free
Maxwell equations.
The function $f<0$ is the square of the timelike
Killing vector field
$\p_t$, $\h$ is the Riemannian 3-metric, and the function
$\psi$ denotes the electrostatic potential.
Majumdar \cite{Majumdar47} proved that if
there is a functional relationship between $f$ and
$\psi$, this must be of the form
\begin{equation}
f=-\psi^2+c\label{Weyl}
\end{equation}
with a constant $c$, generalizing Weyl's result \cite{Weyl17} 
in the case of axial symmetry.
Moreover, he showed that if
\begin{equation}
f=-\psi^2,\label{Majumdar}
\end{equation}
then the 3-metric $h$ is a flat metric, i.e.,
$\h=\d x^2+\d y^2+\d z^2$,
and the field equations reduce to a single Laplace
equation in the 3-dimensional flat space.
Solutions of this type were studied by Papapetrou 
\cite{Papapetrou47} also.
Later, Hartle and Hawking \cite{Hartle&Hawking} investigated
the global structure of the Majumdar-Papapetrou solution,
and they showed that this can be interpreted as describing
an equilibrium state of extremely highly charged black holes.

For a system of  charged dust in Newtonian theory,
a similar equilibrium between gravitational and Coulomb forces
is possible, whenever the mass density $\mu$ and 
charge density $\sigma$ of the charged dust satisfy
\begin{equation}
\sigma=\pm\mu.\label{dust}
\end{equation}
Das \cite{Das62} studied the corresponding
situation in general relativity, and he showed that
the condition (\ref{dust}), as (\ref{Weyl}), implies
the Weyl-Majumdar relation 
(\ref{Majumdar}), and in this case the field equations reduce to
a single non-linear equation; i.e.,
the balance condition (\ref{dust}) makes sense in precisely
the same form in general relativity.
Further discussion concerning the generality of the
assumption (\ref{Weyl}) can be found in 
De and Raychaudhuri. \cite{De&Raychaudhuri68}
Explicit solutions for this system have been found
by many authors, e.g., 
Refs.~\cite{Bonnor65,Kyle&Martin67,Wilson69,Florides77} with spherical symmetry, and
a recent paper \cite{Gurses98} without spatial symmetry.

%In this paper, we shall study the static field of a charged perfect fluid.
%Gautreau and Hoffman \cite{Gautreau&Hoffman73} generalized Das's result
%to this case and showed that the equation (\ref{Weyl}) implies
%\begin{equation}
%\sigma=\pm(\mu+3p)\left(1+\frac{c}{\psi^2}\right)^{\half},
%\end{equation}
%where $p$ is the isotropic pressure of the fluid.
%%%%%%%%%%%%%%%%%%%%%%%%
The balance equation and the Weyl-Majumdar relationship are closely
related, and moreover, it seems that the force balance condition greatly
simplifies the Einstein equation.
The most typical examples are the Weyl and Majumdar-Papapetrou metrics 
mentioned above,
and the G\"urses metric with static charged-dust source. \cite{Gurses98}
In Newtonian theory, 
the gravitational and electrostatic forces acting
on an element of charged dust must be parallel.
This is also the case in Einstein theory; namely,
the Bianchi identity (Euler equation) implies that there is
a functional relationship between the gravitational and 
electrostatic potentials \cite{De&Raychaudhuri68}
(not necessarily in the form of (\ref{Weyl}) or (\ref{Majumdar})).
Such an analogy between the Newtonian and Einstein theory
might hold even in the presence of a third force, e.g.
a magnetic field or a pressure gradient.
A configuration of a magnetic field parallel to the electric field is easily
obtained by a duality rotation of pure electric field. However, there 
will be a magnetic monopole in this case. 
The presence of non-parallel magnetic and electric fields generates a
Poynting flux, contributing to the energy-flux component of the stress-energy 
tensor, which is a characteristic of a stationary field rather than static one.
On the other hand, the (isotropic) matter pressure itself does not generate any 
energy (heat) fluxes. Rather, it effectively contributes only to 
the energy-density and momentum-flux parts of 
the stress-energy tensor by definition, and
one may expect that there is a large class of static 
configurations generated by
a charged perfect-fluid. Such a configuration may be 
considered as the interior metric
of multiple Majumdar-Papapetrou black-holes. Since the 
Majumdar-Papapetrou metrics describe
multiple (extremely highly) charged black-holes, they inevitably have 
timelike singularities in
each horizon, so that consideration of the regular interior 
metric might be 
interesting.
Gautreau and Hoffman \cite{Gautreau&Hoffman73} 
generalized the balance equation in the presence of matter
pressure, assuming Eq.~(\ref{Weyl}), and showed
\begin{equation}
\sigma=\pm(\mu+3p)\left(1+\frac{c}{\psi^2}\right)^{\half},
\end{equation}
where $p$ is the isotropic pressure of the fluid,
 though they did not consider the rest
of the field equations. 
The existence of such a simple balance equation seems to imply
that there are many static configurations.
In fact, Guilfoyle \cite{Guilfoyle99} recently studied this system of a 
charged perfect fluid subject to (\ref{Weyl}) and obtained a large class
of algebraically special (type D) metrics.

In this paper, we investigate a system of a
pure-electrically charged perfect fluid with the Weyl-Majumdar relation from a general
point of view,
and we show that there exists a large class of exact solutions, which can be
algebraically general (type I) and need not have any spatial symmetries.
%%%%%%%%%%%%%%%%%%%%%%%%
As shown below, if the Weyl-Majumdar relation 
(\ref{Majumdar}) is assumed, then the 3-metric $h$ must be that
of a space of constant curvature, and the field equation reduces to
a single non-linear equation.
Moreover, if the fluid obeys a linear equation of state
($p={\rm const.}\times\mu$),
then the field equation becomes a Helmholtz equation on the space of 
constant curvature (\S\ref{2}).
Therefore, solutions are expressed in terms of eigenfunctions of 
the Laplace operator. Some explicit solutions are given in 
\S\ref{3}.

\section{Field equations}\label{2}
The metric of static spacetimes can be expressed in the form
\begin{equation}
\g=-\frac{1}{\phi^2}\d t\o\d t+\phi^2\h,
\end{equation}
where neither the function $\phi>0$ nor the Riemannian 
3-metric $\h$ depends on
the coordinate $t$; i.e., the vector field $\xi=\p_t$ satisfies 
the Killing equation
\begin{equation}
{\cal L}_{\xi}\g=0.\label{Killing}
\end{equation}
The Killing vector field $\xi$ 
of a static spacetime is hypersurface orthogonal, or twist-free,
and thus it satisfies the Frobenius condition
\begin{equation}
\xi\w\d\xi=0,\label{Frobenius}
\end{equation}
where $\xi=-\phi^{-2}\d t$ is the Killing
1-form. (Here and in what follows, we use the same symbol for
vector fields and their corresponding 1-forms.)

The time-space components of the Ricci tensor are obtained from the
equation (see e.g., Ref.~\cite{Carter})
\begin{equation}
16\pi\tilde{\j}=-\ast\d\ast\d\xi,
\end{equation}
where
\begin{equation}
8\pi\tilde{\j}=-\ric(\xi,\p_{\mu})\d x^{\mu},
\end{equation}
and $\ast$ represents the Hodge operator.
The Lichnerowicz theorem states that $\tilde{\j}$ is proportional
to $\xi$ in the static case. \cite{Carter} This can be shown by
\begin{eqnarray}
0&=&\d\ast(\xi\w\d\xi)
=\d\iota_{\xi}\ast\d\xi\nonumber\\
&=&({\cal L}_{\xi}-\iota_{\xi}\d)\ast\d\xi
=16\pi\iota_{\xi}\ast\tilde{\j}\nonumber\\
&=&-16\pi\xi\w\tilde{\j},\label{Lichnerowicz}
\end{eqnarray}
where $\iota_{\xi}$ denotes the interior product with respect to
$\xi$, and the identity ${\cal L}_{\xi}=\d\iota_{\xi}+\iota_{\xi}\d$
has been used.

The time-time component of the Ricci tensor is obtained by calculating
the Laplacian of $f=-\phi^{-2}$:
\begin{eqnarray}
\d\ast\d f&=&\d\ast\d\ix\xi=\d\ast(\Lx-\ix\d)\xi\nonumber\\
&=&\d(\xi\w\ast\d\xi)=\ast(\d\xi:\d\xi)-\xi\w\d\ast\d\xi\nonumber\\
&=&\ast\frac{\gu(\d f,\d f)}{f}+16\pi\ast\gu(\xi,\tilde{\j}).
\end{eqnarray}
Here the colon denotes the inner product of differential forms (i.e.,
$\alpha:\beta=(1/p!)\alpha_{\mu_1\mu_2\cdots\mu_p}
\beta^{\mu_1\mu_2\cdots\mu_p}$ for $p$-forms $\alpha$ and $\beta$).
Therefore, we have
\begin{equation}
8\pi\tilde{\rho}=\ric(\xi,\xi)
%=\ast\d\ast\d f+\frac{\gu(\d f,\d f)}{f}.
=\phi^{-3}*\d*\d \phi+\phi^{-4}\gu(\d \phi,\d \phi).
\end{equation}
The Ricci tensor of the 3-metric $\h$ 
encodes the remaining components 
of the Ricci tensor, and with a direct calculation,
we obtain the expression \cite{Geroch}
\begin{equation}
\ric[\h]=\ric-8\pi \phi^2(\tilde{\rho}\g+2\xi\o\tilde{\j})
+2\frac{\d \phi\o\d \phi}{\phi^2}.
\label{3-ric}
\end{equation}

Next, consider the static
electro-magnetic field $\F=\d \A$, $\Lx\F=0$.
Here we assume that only the electric field exists for static
observers, which implies
\begin{equation}
\xi\w \F=0.
\end{equation}
%%%%%%%%%%%%%%%%%%%
In the source-free Einstein-Maxwell static case,
 this assumption is justified (up to duality rotation)
 by the Einstein equation,
while in the charged perfect fluid case, this condition
is equivalent to imposing the condition that the 
fluid element as well as the gravitational field is static. Moreover,
there is no freedom of the duality rotation in this case.
%%%%%%%%%%%%%%%%%%%
Here, the Maxwell equation $\d\F=0$ becomes
\begin{equation}
\d\ix\F=(\Lx-\ix\d)\F=0,\label{curlE}
\end{equation}
while the other Maxwell equation, $\d\ast\F=4\pi\sigma*\u$, reduces to
\begin{eqnarray}
\d\ast\ix\F&=&-\d(\xi\w\ast\F)=
-\ast(\d\xi:\F)+4\pi\sigma\ast\gu(\xi,\u)\nonumber\\
&=&-2\ast\frac{\gu(\ix\F,\d \phi)}{\phi}
+4\pi\sigma\ast\gu(\xi,\u),\label{divE}
\end{eqnarray}
where $\u$ and $\sigma$ denote the velocity and the charge density
of the fluid, respectively. 
Equation (\ref{curlE}) implies that locally there exists a real
function (electrostatic potential) $\psi$ such that $\d\psi=\ix\F$ holds.
Then the full Maxwell equations reduce to a single Poisson equation,
\begin{equation}
\ast\d\ast\d\psi=2\frac{\gu(\d\psi,\d \phi)}{\phi}-4\pi\sigma\gu(\xi,\u).
\label{Poisson}
\end{equation}
On the other hand, the stress-energy tensor of the Maxwell field $\T_F$ is
\begin{equation}
8\pi\T_F(\p_{\mu},\p_{\nu})=\gu(\iota_{\p_{\mu}}\F,\iota_{\p_{\nu}}\F)
+\gu(\iota_{\p_{\mu}}*\F,\iota_{\p_{\nu}}*\F)
\label{Tf}
\end{equation}
In pure electric cases, $\j_F=-\T_F(\xi,\p_{\mu})\d x^{\mu}$
is proportional to $\xi$. \cite{Carter} Thus, the energy-flux components
of the stress-energy tensor vanish,
\begin{equation}
\xi\w\j_F=0,
\end{equation}
while the energy-density component becomes
\begin{equation}
8\pi\rho_F=8\pi\T_F(\xi,\xi)=\gu(\d\psi,\d\psi).
\label{Tf1}
\end{equation}
In terms of $\j_F$ and $\rho_F$, Eq.~(\ref{Tf}) can be written as
\begin{equation}
8\pi\T_F=8\pi\phi^2(\rho_F\g+2\xi\o\j_F)-2\phi^2\d\psi\o\d\psi.
\label{Tf2}
\end{equation}

Next, consider a perfect fluid. Its stress-energy tensor has the form
\begin{equation}
\T_m=(\mu+p)\u\o\u+p\g,
\end{equation}
or equivalently
\begin{equation}
\tilde{\T}_m=\T_m-\frac{1}{2}{\rm tr}(\T_m)\g=(\mu+p)\u\o\u
+\frac{1}{2}(\mu-p)\g.
\label{Tm}
\end{equation}
The fluid velocity $\u$ must be proportional to $\xi$; i.e.,
$\u=\phi\xi$. (This is a consequence of the time-space components
of the Einstein equation.)
In analogy to the electric field, we define
\begin{equation}
\tilde{\j}_m=-\tilde{\T}_m(\xi,\p_{\mu})\d x^{\mu}=\frac{1}{2}(\mu+3p)\xi
\end{equation}
and
\begin{equation}
\tilde{\rho}_m=\tilde{\T}_m(\xi,\xi)=\frac{(\mu+3p)}{2\phi^2}.
\end{equation}
The equation of hydrostatic equilibrium (Euler equation)
 is obtained from the Bianchi identity 
for the stress-energy tensor 
$\nabla\cdot(\T_F+\T_m)=0$, which gives 
\begin{equation}
\d\ln\phi=\frac{1}{\mu+p}(\d p+\sigma\phi\d\psi).\label{euler}
\end{equation}

Now we consider the Einstein equation
\begin{equation}
\ric=8\pi(\T_F+\tilde{\T}_m).
\end{equation}
The time-space component equations of the field equation
$\xi\w(\tilde{\j}-\j_F-\tilde{\j}_m)=0$ are identically satisfied, 
while the time-time component
$\tilde{\rho}=\rho_F+\tilde{\rho}_m$ leads to
\begin{equation}
\ast\d*\d\phi=-\frac{\gu(\d\phi,\d\phi)}{\phi}
+\phi^3\gu(\d\psi,\d\psi)+4\pi\phi(\mu+3p).
\label{Hamiltonian}
\end{equation}
From Eqs.~(\ref{3-ric}), (\ref{Tf2}) and
(\ref{Tm}), we obtain the remaining equation,
\begin{eqnarray}
\ric[\h]+16\pi\phi^2p\h
=2\phi^{-2}\d\phi\o\d\phi-2\phi^2\d\psi\o\d\psi.
\label{spatial}
\end{eqnarray}
%the equations (\ref{Poisson}),~(\ref{Hamiltonian}) and 
%(\ref{spatial}) form a complete set of the field equations.
Our task is to solve 
Eqs.~(\ref{Poisson}),~(\ref{euler}),~(\ref{Hamiltonian})
and (\ref{spatial}). However, we need additional
conditions, such as an equation of state,
to make this system deterministic.

To solve Eq.~(\ref{spatial}), we assume the Weyl-Majumdar relation 
(\ref{Majumdar}), or
\begin{equation}
\psi+\epsilon\phi^{-1}=0,~~~(\epsilon=\pm 1).
\end{equation}
Then the r.h.s. of
Eq.~(\ref{spatial}) vanishes,
and the 3-dimensional manifold with metric $\h$ becomes an Einstein space;
%%%%%%%%%%%%%%%%%%
i.e., the Ricci tensor has only a trace part (scalar curvature).
Since in three or higher dimensional Einstein space, the scalar
curvature must be constant (which can be seen from the contracted
Bianchi identity),
%%%%%%%%%%%%%%%%%%
%which implies that $p$ is proportional to $\phi^{-2}$.
%%%%%%%%%%%%%%%%%%
the matter pressure $p$ is proportional to $\phi^{-2}$
with a constant ratio.
%%%%%%%%%%%%%%%%%%
Without loss of generality, we may take $p=-\kappa/8\pi\phi^2$
($\kappa=0,\pm1$).
%Since a 3-dimensional Einstein space is a space of constant curvature,
%the geometry is characterized by a ,
Since a 3-dimensional Einstein space is a space of constant curvature
(namely, the metric as well as the curvature is charactrized by a constant
scalar curvature), we have
\begin{equation}
\h=\frac{\d x^2+\d y^2+\d z^2}
{\left[1+{\textstyle{1\over 4}}\kappa(x^2+y^2+z^2)\right]^{2}}.
\end{equation}
%\begin{equation}
%\psi+\epsilon\phi^{-1}={\rm const.},~~~(\epsilon=\pm 1).
%\end{equation}
%with a non-zero constant $a$.
Then, Eqs.~(\ref{Poisson}) and
(\ref{Hamiltonian}) reduce to
\begin{equation}
(\Delta+\lambda)\phi=0
\label{Helmholtz}
%+\frac{\mu+3p}{f})
%\frac{1}{\E+1/4a^2}=0
\end{equation}
and
\begin{equation}
\sigma=\epsilon(\mu+3p),\label{balance}
%\frac{a(\E+1/4a^2)}{|a(\E+1/4a^2)|},
\end{equation}
where $\Delta$ denotes the Laplacian with respect to $\h$, and
$\lambda=4\pi\phi^2(\mu+3p)$.
It is easily checked that the Euler equation (\ref{euler}) is
consistent with (\ref{balance}).

Equation~(\ref{Helmholtz}) is non-linear. However, it can be made linear if
 we assume a linear (isothermal-type) equation of
state for the fluid, 
i.e. $\mu={\rm const}\times p (={\rm const}\times\phi^{-2})$. 
Then, $\lambda$ becomes constant, and 
Eq.~(\ref{Helmholtz}) is simply a Helmholtz equation in the space of 
constant curvature, whose solutions are known (see, e.g., Ref.~\cite{Harrison}).

\section{Examples}\label{3}
Here we study some explicit solutions of Eq.~(\ref{Helmholtz}).
We mainly consider the spherically symmetric solutions
 as  simple examples.
We assume the dominant energy condition,
which requires $\mu\ge |p|$.
\subsection{Positive pressure ($\kappa=-1$)}
In this case, the dominant energy condition requires $\lambda\ge 2$.
In the spherical polar coordinate system $\{r,\vartheta,\varphi\}$,
the 3-metric $\h$ can be written as
\begin{equation}
\h=\eta(r)^2[\d r^2+r^2(\d\vartheta^2+\sin^2\vartheta\d\varphi^2)],
\end{equation}
where $\eta=(1-r^2/4)^{-1}$ $(0<r<2)$. 
Equation (\ref{Helmholtz}) for the
function $\phi(r)$ becomes
\begin{equation}
\frac{1}{\eta^3 r^2}\frac{\d}{\d r}\eta r^2\frac{\d}{\d r}\phi
+\lambda\phi=0.
\label{radial}
\end{equation}
The general solution of this equation is
\begin{equation}
\phi=l\frac{1-r^2/4}{r}\sin
\left[ (\lambda-1)^{\half}\ln\frac{2+r}{2-r}+\delta\right] ,
\label{pphi}
\end{equation}
where $l>0$ and $0\le\delta\le\pi$ are integration constants.
In general, the set on which
$\phi=0$ is a singularity, since, for example, the field
invariant $\F:\F=-\phi^{-2}\gu(\d\phi,\d\phi)$ blows up there.
%(only the exception is $\delta=0$, $r=0$, in which case
%the metric comes close to flat near the origin $r=0$),
Hence this solution must have a singularity, and the situation
is the same even without spherical symmetry.
The locations of the singularities are given by
\begin{equation}
r=2\tanh\frac{(n\pi-\delta)}{2(\lambda-1)^{\half}},~~ 
(n=0,\pm1,\pm2,\cdots),
\end{equation}
at which the area radius $\R=|\phi\eta r|$ vanishes,
so that these singularities are actually point-like, and each connected
component of the spacetime has in general the topology $S^3$ minus two points.
(Without spherical symmetry, the singularity might be line-like or surface-like,
and the spacetime topology would be $S^3$ minus several such sets.)

Next, let us consider the solution (\ref{pphi}) in the neighborhood of $r=0$.
We may assume that 
%%%%%%%%%%%%%%%
$0\le\delta\le\pi/2$, 
%%%%%%%%%%%%%%%%%%%%%%%%%%%%%%%%%%
since the transformation
$r\mapsto-r$ corresponds to $\delta\mapsto\pi-\delta$.
Then, let us consider the region $r<0$.
Whenever $\delta\ne0$, this solution diverges at $r=0$ like $l\sin\delta/r$.
However, this is only a coordinate singularity.
The hypersurface $r=0$ is a null hypersurface, on which
the Killing vector becomes null, and it is generated
by spheres of radius $\R=l\sin\sigma$. In fact, whenever 
%%%%%%%%%%%%%
$\delta\ne 0, \pi/2$,
%%%%%%%%%%%%%%%%%%%%%%%%%%
the region $r<0$ is
extended analytically to the region $r>0$ across this null hypersurface.
To see this, let us consider the coordinate transformation
\begin{equation}
v=t+\int^r\phi^2\eta\d r.
\end{equation}
Then the metric transforms to
\begin{eqnarray}
&&\g=-\frac{\d v^2}{\phi^2}+\frac{2}{1-r^2/4}\d v\d r\nonumber\\
&&{}+l^2\sin^2\left[(\lambda-1)^{\half}\ln\frac{2+r}{2-r}\right]
(\d\vartheta^2+\sin^2\vartheta\d\varphi^2),
\end{eqnarray}
which is manifestly non-singular at $r=0$.

Moreover, we can consider a class $C^1$ extension of 
the region $r<0$ to the exterior field of a black-hole spacetime.
The Misner-Sharp mass $m_H$ 
(which coincides with the Hawking mass or Kodama mass in the case of spherical symmetry)
with respect to the surface $r=0$ is
\begin{equation}
m_H=\frac{l\sin\delta}{2}.
\end{equation}
On the other hand, the electric charge inside a two-surface $S$ can be
evaluated as
\begin{equation}
q=-\frac{1}{4\pi}\int_S*\F.
\end{equation}
In particular, when a surface defined by
$r={\rm const}$ is chosen as $S$, we find
\begin{equation}
q=\epsilon \frac{r^2}{1-r^2/4} \frac{\d|\phi|}{\d r},
\end{equation}
where `inside' corresponds to smaller $r$, so that the electric charge
inside the surface $r=0$ is
\begin{equation}
q_H=\epsilon l\sin\delta=2\epsilon m_H.
\end{equation}
This relationship is identical to that satisfied by the extreme
Reissner-Nordstr\"{o}m spacetime on the horizon.
Furthermore, the mass density and matter pressure of the fluid vanish on
the surface $r=0$.
These evaluations suggest that the region $r<0$ with the 
metric characterized by the Eq.~(\ref{pphi}) can be matched at $r=0$
to the exterior field with the extreme Reissner-Nordstr\"{o}m metric.
By  rescaling  the coordinates according to
\begin{equation}
\tr=l(\lambda-1)^{\half}\cos\delta~r,~~~
\tt=\frac{t}{l(\lambda-1)^{\half}\cos\delta},
\end{equation}
the metric transforms to
\begin{equation}
\g=-\frac{\d \tt^2}{\tphi^2}+\tphi^2\teta^2
\left[\d\tr^2+\tr^2(\d\vartheta^2+\sin^2\vartheta\d\varphi^2)\right],
\label{ng}
\end{equation}
where
\begin{equation}
\tphi=\frac{l}{\tr\teta}\sin\left[(\lambda-1)^{\half}
\ln\frac{2l(\lambda-1)^{\half}\cos\delta+\tr}
{2l(\lambda-1)^{\half}\cos\delta-\tr}+\delta\right],
\end{equation}
and
\begin{equation}
\teta=\left[1-\frac{\tr}{4\l^2(\lambda-1)\cos^2\delta}\right]^{-1}.
\end{equation}
Note that $\tphi$ behaves near $\tr=0$ like
\begin{equation}
\tphi=1+\frac{l\sin\delta}{\tr}+O(\tr^2).
\label{pexpansion}
\end{equation}
Next, we introduce the advanced time coordinate
\begin{equation}
v=\tt+\int^{\tr}\tphi^2\teta\d\tr.
\end{equation}
Then the metric (\ref{ng}) becomes
\begin{equation}
\g=-\frac{\d v^2}{\tphi^2}+2\teta\d v\d\tr
+\tphi^2\teta^2\tr^2(\d\vartheta^2+\sin^2\theta\d\varphi^2),
\label{pinterior}
\end{equation}
while the metric of the extreme Reissner-Nordstr\"{o}m has the form
\begin{eqnarray}
\g_{RN}&=&-\left(1+\frac{m}{\tr}\right)^{-2}\d v^2+2\d v\d\tr
\nonumber\\
&&{}+\left(1+\frac{m}{\tr}\right)^2\tr^2
(\d \vartheta^2+\sin^2\vartheta\d\varphi^2).
\label{RN}
\end{eqnarray}
If the mass parameter (ADM mass) of the metric (\ref{RN}) is
\begin{equation}
m=l\sin\delta,
\end{equation}
then these two metrics are matched on the hypersurface $\tilde{r}=0$, and it
is easily seen from (\ref{pexpansion}) that this extension is class $C^1$.
In this case, the exterior metric (\ref{RN}) and
the interior metric (\ref{pinterior})  correspond to $\tr>0$ and 
$\tr<0$, respectively. 
We can similarly verify that the region
$r>0$ for (\ref{pphi}) 
can be matched to the $m=l\sin\delta$ interior metric 
(i.e., the metric (\ref{RN}) for $m=-l\sin\delta<0$, $\tr>0$), 
and if $l\sin\delta<0$, then the regions $r<0$ and $r>0$ for (\ref{pphi}) 
can be matched to the  interior ($m=-|l\sin\delta|$) 
and exterior ($m=|l\sin\delta|$) metric (\ref{RN}),
respectively.

Of course, we may also obtain solutions without spherical symmetry
using the method of multipole expansion.
Instead, here we give an example by superposing
monopole solutions with different centers $(x_i,y_i,z_i)$:
\begin{equation}
\phi=\sum_i
l_i\frac{1-r_i^2/4}{r_i}
\sin\left[ 
(\lambda-1)^{\half}\ln\frac{2+r_i}{2-r_i}
+\delta_i
\right],
\label{psuperposition}
\end{equation}
where
\begin{equation}
r_i=[(x-x_i)^2+(y-y_i)^2+(y-y_i)^2]^{\half}.
\end{equation}
Here, the quantities $l_i$ and $\delta_i$ are constants. Each hole $r_i=0$ can be
matched to the Reissner-Nordstr\"{o}m metric with $m=l_i\sin\delta_i$.
%%%%%%%%%%%%%%%
%%%%%%%%%%%%%%%
\subsection{Dust $\kappa=0$}
For completeness, we briefly discuss this special case.
We only give some explicit
solutions here, since the properties of these solutions are quite similar to those in the
positive pressure case.
Recently, G\"{u}rses \cite{Gurses98} also found this class.
The field equation is the ordinary Helmholtz equation,
\begin{equation}
\left(\frac{\p^2}{\p x^2}+\frac{\p^2}{\p y^2}+\frac{\p^2}{\p z^2}+\lambda
\right)\phi=0,
\end{equation}
and the energy condition requires $\lambda\ge 0$.
The general spherically symmetric solution is
\begin{equation}
\phi=l\frac{\sin(\lambda^{\half}r+\delta)}{r},
\label{dphi}
\end{equation}
where $l>0$ and $0\le\delta\le\pi$ are integration constants.
As in the positive pressure case, the solution 
vibrates around $\phi=0$ in general, so that there always exists a
singularity. The solution is matched to the exterior 
Reissner-Nordstr\"{o}m metric with $m=l\sin\delta$ at $r=0$ with
class $C^1$ differentiability. Moreover, in the case of 
spherically symmetric static dust solutions, the metric can be matched 
to an extreme Reissner-Nordstr\"{o}m metric 
on any symmetric sphere by virtue of the absence of matter pressure.

The solution without spatial symmetry can be expressed in terms of
the spherical harmonics and the spherical Bessel functions.
We simply give a multi-center solution 
\begin{equation}
\phi=\sum_il_i\frac{\sin(\lambda^{\half}r_i+\delta)}{r_i}.
\label{dsuperposition}
\end{equation}
For more information, see Ref.~\cite{Gurses98}.
%%%%%%%%%%%%%%%
%%%%%%%%%%%%%%%
\subsection{Negative pressure ($\kappa=1$)}
In this case the dominant energy condition requires $\lambda\ge-1$.
Consider the solution of Eq.~(\ref{radial}) with 
$\eta=(1+r^2/4)^{-1}$.
The general solution is
\begin{equation}
\phi=l\frac{1+r^2/4}{r}\sin\left[(1+\lambda)^{\half}
\arcsin\frac{r}{1+r^2/4}+\delta\right],
\label{nphi}
\end{equation}
where $l>0$ and $0\le\delta\le\pi$ are integration constants, and 
the $\arcsin$ above should be regarded as
\begin{equation}
\arcsin \frac{r}{1+r^2/4}= \left\{
\matrix{
-\pi-\Arcsin\frac{r}{1+r^2/4} &(r<-2)\cr
\Arcsin\frac{r}{1+r^2/4}     &(-2\le r\le 2) \cr
\pi-\Arcsin\frac{r}{1+r^2/4} &(r>2)
}
\right.
\end{equation}
in terms of the $\Arcsin$ (which takes the principal value).
Unlike the case of positive pressure or dust, these solutions need not have
any singularity in the appropriate range of $\lambda$. In fact,
if $-1\le\lambda\le-3/4$ ($|p|\le\mu\le 3/2|p|$) and
$(1+\lambda)^{\half}\pi\le\delta\le\pi-(1+\lambda)^{\half}\pi$,
then $\phi$ gives a regular solution for all $r$, 
and
if $\delta=0$ and $-1\le\lambda\le 0$ ($|p|\le\mu\le 3|p|$),
then it gives a regular solution for $r\ge 0$. 

As in the other cases, we can match the
solution to the exterior metric ($\tr>0$) 
of the extreme Reissner-Nordstr\"{o}m
spacetime (\ref{RN}) at $r=0$ whenever $\delta\ne 0,\pi/2,\pi$.
Note that we may assume $0<\delta<\pi/2$, since
$\delta$ is transformed to $\pi-\delta$ by $r\mapsto-r$.
Let us consider the coordinate transformation
\begin{equation}
\tr=l(1+\lambda)^{\half}\cos\delta~r,
~~~\tt=\frac{t}{l(1+\lambda)^{\half}\cos\delta}.
\end{equation}
Then the metric can be written
\begin{equation}
\g=-\frac{\d\tt^2}{\tphi^2}+\tphi^2\teta^2
\left[\d\tr^2+\tr^2(\d\vartheta^2+\sin^2\vartheta\d\varphi^2)\right],
\end{equation}
where
\begin{equation}
\tphi=\frac{l}{\tr\teta}\sin\left[
(1+\lambda)^{\half}\arcsin
\frac{\tr\teta}{l(1+\lambda)^{\half}\cos\delta}+\delta\right],
\end{equation}
and
\begin{equation}
\teta=\left[1+\frac{\tr^2}{4l^2(1+\lambda)\cos^2\delta}\right]^{-1}.
\end{equation}
Furthermore, the transformation
\begin{equation}
v=\tt+\int^{\tr}\tphi^2\teta\d\tr
\end{equation}
gives a metric in the form
\begin{equation}
\g=-\tphi^{-2}\d v^2+2\teta\d v\d\tr
+\tphi^2\teta^2\tr^2(\d\vartheta^2+\sin^2\vartheta\d\varphi^2).
\label{ninterior}
\end{equation}
The region $\tr<0$ with this metric can be matched  
to the region $r>0$ with the metric (\ref{RN}) with
\begin{equation}
m=l\sin\delta.
\end{equation}
In this case also, the extension is class $C^1$, since
\begin{equation}
\tphi=1+\frac{l\sin\delta}{\tr}+O(\tr^2),~~~
\teta=1+O(\tr^2).
\end{equation}

Unlike the case of a positive pressure fluid, the metric (\ref{nphi}) or
(\ref{ninterior}) can be
regular for all $r<0$ ($\tr<0$)
if $\delta\ge(1+\lambda)^{\half}\pi$ is satisfied.
In this case, the surface $r=-\infty$ is neither infinity nor
a singularity. In fact, it can be shown that this surface is a null
 hypersurface and that the spacetime can be extend beyond $r=-\infty$,
unless $\delta=(1+\lambda)^{\half}\pi$, in which case $r=-\infty$
simply corresponds to a regular center.
To see this, consider the inverse of the radial coordinate,
\begin{equation}
\br=4/r.
\end{equation}
In terms of this coordinate, the metric transforms to
\begin{equation}
\g=-\frac{\d t^2}{\bphi^2}+\bphi^2\bfeta^2
\left[\d\br^2+\br^2(\d\vartheta^2+\sin^2\vartheta\d\varphi^2)\right],
\end{equation}
where
\begin{eqnarray}
\bphi=l\frac{1+\br^2/4}{\br}\sin\left[
(1+\lambda)^{\half}\arcsin\frac{\br}{1+\br^2/4}\right.\nonumber\\
\left.+\pi+(1+\lambda)^{\half}\pi-\delta\right],
\end{eqnarray}
and
\begin{equation}
\bfeta=\left(1+\quota\br^2\right)^{-1}.
\end{equation}
Hence, the effect of this transformation is simply
\begin{equation}
\delta\mapsto\pi+(1+\lambda)^{\half}\pi-\delta,
\end{equation}
so that the metric can be analytically extended across the null hypersurface
$\br=0$ in a manner similar explained above, and  it can also be matched
to the interior region ($\tr<0$) 
with the Reissner-Nordstr\"{o}m metric
(\ref{RN}) with 
$m=l\sin[\delta-(1+\lambda)^{\half}\pi]$.

Finally, we obtain a multi-center solution in an obvious way: 
\begin{equation}
\phi=\sum_il_i\frac{1+r_i^2/4}{r_i}\sin\left[(1+\lambda)^{\half}
\arcsin\frac{r_i}{1+r_i^2/4}+\delta_i\right].
\label{nsuperposition}
\end{equation}

\section{Conclusions}
We have investigated the static field of a charged perfect
fluid, and shown that only two assumptions,
the Weyl-Majumdar relation (\ref{Majumdar}) and a
linear equation of state, allow us to reduce the field equations
to a single linear equation.
Thus, any solution of this type can be described by
an eigenfunction of the Laplacian in the 3-dimensional
space of constant curvature, and the eigenvalue determines 
the equation of state,
i.e., the ratio of the energy density to pressure.
These solutions have some desirable properties. For example,
they obey a simple equation of state, as required,
neither the  mass density nor the matter pressure changes sign,
and they need not have any spatial symmetry.

The mechanism responsible for this simplification of field equations
may be understood in terms of
force balance among the gravitational force, the Coulomb force and
the pressure gradient.
The function $\phi^{-2}$ is the norm of the static Killing vector,
and thus it can be interpreted as the gravitational potential.
Then Eq.~(\ref{Majumdar}) gives a functional relationship
between the gravitational and electrostatic potential, which implies
that the gravitational and Coulomb forces are parallel.
This property characterizes the Weyl\cite{Weyl17} and 
Majumdar-Papapetrou\cite{Majumdar47,Papapetrou47} solutions.
%and explains why they permitts a large class of solutions without spherical symmetry.
One main point of this paper is that the pressure gradient automatically 
becomes parallel to both the gravitational and Coulomb forces 
as seen from the Einstein
equation, even in the presence of a perfect fluid, which is implicit in the
spherically symmetric case.
Thus, the balance equation among the three relevant forces, which is
originally a vector equation, becomes a scalar equation (\ref{balance}),
explaining why a large class of inhomogeneous metrics are obtained.
%%%%%%%%%%%%%%%%%%%%
Though we have discussed only the situatioin in which
 equipotential surfaces
of the gravitational potential, the electrostatic potential,
and the matter pressure coincide, it is not a
necessary condition for the static configuration;
there would be a situation in which
 the vector balance equation holds at each point. However, 
such a spacetime might have some spatial symmetry, such as
axial or cylindrical symmetry, or the equation of state might
depend on spacetime points.
The static configuration considered here may 
be regarded as a final state of a non-rotating object composed of charged fluid.
Through the realistic contraction of such an object with a given equation of state,
the scalar balance equation might be satisfied in the final equilibrium state,
even if it is not spherically symmetric.
%%%%%%%%%%%%%%%%%%%%

Another characterization of the solution can be given in terms of the
Ernst potentials\cite{Ernst}.
In analogy to the stationary Einstein-Mawell system, we can
define the Ernst potential by ${\cal E}=1/\phi^2-\psi^2$. Then
Eq.~(\ref{Hamiltonian}) can be replaced by the
Poisson equations for ${\cal E}$. However, the assumption
(\ref{Majumdar}) implies that ${\cal E}=0$, so that this equation reduces
to an algebraic equation, which is just the balance equation 
(\ref{balance}).

We have also shown that  solutions corresponding to 
Eqs.~(\ref{pphi}), (\ref{dphi}) and (\ref{nphi})
 can be matched to the extreme
Reissner-Nordstr\"{o}m metric at the horizon. However,
this interior solution has a central singularity unless the matter
pressure is negative.
In the negative pressure case, the interior solution (\ref{nsuperposition})
can be regular for an appropriate range of parameters, and then 
$r\rightarrow-\infty$ does not represent conformal infinity but, rather,
represents
another `hole' connected by another interior metric, or regular center.
Since the matching conditions are determined by the
behavior near the horizon,  solutions
(\ref{nsuperposition}), (\ref{dsuperposition}) and (\ref{psuperposition})
 could be matched to an appropriate
Majumdar-Papapetrou metric \cite{Majumdar47,Papapetrou47,Hartle&Hawking} 
at $r_i=0$ in a manner similar to that described here.

The existence of a regular interior metric has some connection with the
Buchdahl theorem. \cite{buchdahl}
Roughly speaking, the Buchdahl thorem states that a relativistic star
composed of perfect fluid with mass $M$ and radius $R$ can exist
only if $R>(9/4)M$. The charged generalization
of the Buchdahl theorem has been proved by Yu and Liu.\cite{yu}
They showed that the radius of a star of charged fluid has a similar lower bound
if the total charge of the star does not exceed the extreme limit $Q=M$.
In the extreme case, there is a series of interior solutions in which the
surface of the star can be arbitrarily close to a black-hole 
horizon,\cite{defelice95,defelice} showing that $\inf R=M$. On the
other hand, our example shows that there actually exists an $R=M$ star.

\acknowledgements
We would like to acknowledge many helpful discussions with 
Professor~Humitaka Sato, Professor~Ken-ichi Nakao and 
Dr. Akihiro Ishibashi. This work was supported by
research fellowships of the Japan Society for 
the Promotion of Science for Young Scientists.

\end{document}